\documentclass[aps, prb, reprint, superscriptaddress]{revtex4-1}

\RequirePackage[dvips]{graphicx}

\newcommand{\degr}{$^{\mathrm o}$}
\newcommand{\pilatus}{\mbox{PILATUS} \mbox{100K}}
\newcommand{\GenX}{\emph{GenX}}

\begin{document}



\title{Presence of a (1$\times$1) oxygen overlayer on bare ZnO(0001) surfaces and at Schottky interfaces}



\author{Christian M. Schlep{\"u}tz}
\email[]{cschlep@umich.edu}
\affiliation{Department of Physics, University of Michigan, Ann Arbor, MI 48109, USA}

\author{Yongsoo Yang}
\affiliation{Department of Physics, University of Michigan, Ann Arbor, MI 48109, USA}

\author{Naji S. Husseini}
\affiliation{Department of Physics, University of Michigan, Ann Arbor, MI 48109, USA}

\author{Robert Heinhold}
\affiliation{MacDiarmid Institute for Advanced Materials and Nanotechnology, New Zealand}
\affiliation{Department of Electrical and Computer Engineering, University of Canterbury, Christchurch 8140, New Zealand}

\author{Hyung-Suk Kim}
\affiliation{MacDiarmid Institute for Advanced Materials and Nanotechnology, New Zealand}
\affiliation{Department of Electrical and Computer Engineering, University of Canterbury, Christchurch 8140, New Zealand}

\author{Martin W. Allen}
\affiliation{MacDiarmid Institute for Advanced Materials and Nanotechnology, New Zealand}
\affiliation{Department of Electrical and Computer Engineering, University of Canterbury, Christchurch 8140, New Zealand}

\author{Steven M. Durbin}
\affiliation{MacDiarmid Institute for Advanced Materials and Nanotechnology, New Zealand}
\affiliation{Department of Electrical Engineering and Department of Physics, University at Buffalo, Buffalo, NY 14260, USA}

\author{Roy Clarke}
\affiliation{Department of Physics, University of Michigan, Ann Arbor, MI 48109, USA}


\date{\today}

\begin{abstract}
The atomic surface and interface structure of bare and metal-coated ZnO(0001) Zn-polar wafers were investigated via surface x-ray diffraction. All bare samples showed the presence of a (1$\times$1) overlayer of oxygen atoms located at the on-top position above the terminating Zn atom, a structure predicted to be unstable by several density functional theory calculations. The same oxygen overlayer is clearly seen at the interface of ZnO with both elemental and oxidized metal contact layers. No significant atomic relaxations are observed at surfaces and interfaces processed under typical device fabrication conditions.
\end{abstract}



\maketitle 



ZnO has many important technological applications in fields as diverse as catalysis, gas sensing, corrosion prevention, and optoelectronics.\cite{Woll:2007p2914} A precise knowledge of the structural properties of the surfaces and interfaces involved in these applications is a prerequisite for their control and optimization. However, despite a large number of experimental and theoretical studies, the surface and interface structure of ZnO remain a topic of intense debate, with many alternative structures having been described.\cite{Jedrecy:2000p2901,Wander:2001p2905,Dulub:2003p2908,Woll:2007p2914,King:2008p2894,Valtiner:2010p3053,Kresse:2003p2990,Valtiner:2009p2890,Valtiner:2010p3048} This lack of consensus arises primarily from the competition of various surface stabilization mechanisms and a host of different preparation procedures and ambient conditions, resulting in a complex phase diagram.\cite{Kresse:2003p2990,Valtiner:2009p2890,Valtiner:2010p3048}

In this letter, we report on a systematic study of the atomic structure of bare ZnO surfaces and metal Schottky contacts to ZnO prepared under typical device fabrication conditions. Schottky contacts are important building blocks in many electronic devices, and an understanding of their interface structure is important, since electronic and structural properties are usually strongly correlated. This is particularly true for semiconductors characterized by highly ionic bonding (e.g., ZnO), in which small atomic displacements can potentially result in large changes in electronic behavior.

\begin{figure}
  \includegraphics[width=1.0\columnwidth]{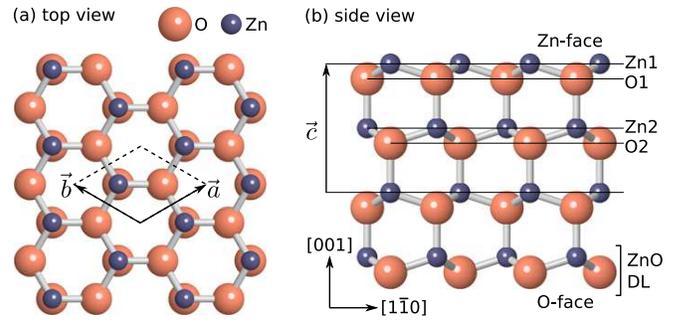}
  \caption{\label{fig:ZnO_structure}(color online) Atomic structure of a bulk-like ZnO(0001) surface. (a) top view with the surface unit cell. (b) side view along the [110] direction. Note the presence of two ZnO double layers (DLs) within the unit-cell height $c$.}
\end{figure}


Single-crystal ZnO$(0001)$ wafers were hydrothermally grown along the +c-axis by Tokyo Denpa Co. Ltd. and epi-polished to low miscut angles ($<0.1$\degr).\cite{Ohshima:2004p3113} For measurements on bare surfaces, samples no. 1-3 were ultrasonically cleaned using organic solvents, dried in N$_2$, and subsequently exposed to air. Polycrystalline Schottky contacts were deposited on other samples after the same cleaning procedure, either as plain gold by thermal evaporation (samples no. 4,5) or non-stoichiometric iridium oxide (IrO$_{x}$) layers by eclipse pulsed laser deposition in an oxygen ambient (samples no. 6,7). The latter method has been shown to consistently produce high quality Schottky contacts to ZnO.\cite{Allen:2009p2952}


Surface x-ray diffraction (SXRD) offers the unique capability to access the buried interface structures of the Schottky contacts non-destructively, and it provides picometer accuracy for the determination of atomic positions. SXRD experiments were carried out on the Materials Science beamline X04SA at the Swiss Light Source (SLS) and on the GeoSoilEnviroCARS beamline, 13-BMC, at the Advanced Photon Source (APS) using 12.398 keV and 15.000 keV photons, respectively. A \pilatus\ pixel detector\cite{Schleputz:2005p2,Kraft:2009p2766} was used for fast and reliable data acquisition.

In-plane line scans along high-symmetry directions gave no evidence for surface or interface reconstructions on any sample. The out-of-plane crystal truncation rod (CTR) measurements revealed a 6-fold rotational symmetry of the diffraction pattern ($p6mm$), even though a perfectly bulk-truncated ZnO crystal is expected to have $p3m1$ symmetry. However, there are two ZnO double layers (DLs) in the ZnO unit cell (Fig.~\ref{fig:ZnO_structure}), both of which result in a chemically identical termination layer but a 180\degr\ rotation of the $p3m1$ diffraction pattern. The overall $p6mm$ symmetry is therefore consistent with the presence of 1/2 unit-cell height terrace steps and equal proportions of both terminations on the surface. Atomic force microscopy measurements on hydrothermal ZnO$(0001)$ Zn-polar surfaces cleaned in the same way have previously showed evidence of triangular islands and pits with 180\degr\ rotation between triangles on terraces separated by a single DL step.\cite{Allen:2009p2952}

For each sample, at least 8 symmetrically inequivalent ($p6mm$) CTRs plus several equivalent rods were recorded, typically resulting in 400-700 averaged structure factors (SFs) per data set and systematic errors of 5-10\% between symmetry equivalents. A specialized module to model SXRD data for the genetic algorithm refinement program \GenX\cite{Bjorck:2007p1384} was used to fit the measured SFs. Utilizing differential evolution algorithms, \GenX\ efficiently avoids becoming trapped in local minima. To verify the reproducibility and uniqueness of a solution, all fits were repeated at least ten times with random parameter initializations. To give the low-intensity regions of the CTRs a similar weight in the fit as the high-intensity points close to the Bragg peaks, a logarithmic $R$-factor ($R_{\mathrm{log}}$) was employed for the fitting figure of merit (FOM).
\footnote{
  $R_{\mathrm{log}}=
  \sum_i{\left|\log_{10}(\left|F_i^{\mathrm{exp}}
  \right|)-\log_{10}(\left|F_i^{\mathrm{calc}}\right|)\right|} /
  \sum_i{\log_{10}(\left|F_i^{\mathrm{exp}}\right|)}$

}
All final fit results are also given in terms of the standard crystallographic $R$-factor.
\footnote{
  $R=\sum_i{\left|\left|F_i^{\mathrm{exp}}\right|-
  \left|F_i^{\mathrm{calc}}\right|\right|} /
  \sum_i{\left|F_i^{\mathrm{exp}}\right|}$ 
}
 
To identify the characteristic structural features, a large number of different models with varying degrees of complexity based on the bulk-like surface structure shown in Fig.~\ref{fig:ZnO_structure} were tested initially. Atomic z-displacements ($\Delta z$), occupations ($p_{\mathrm{occ}}$), and Debye-Waller (DW) factors of all atoms in up to four ZnO DLs were allowed to vary freely. The local $p3m1$ surface symmetry only permits atomic movements $\Delta z$ along the z-direction and allows atoms to be present in just three different positions in the unit cell: at $(x,y) = (0, 0)$ (fcc hollow), $(1/3, 2/3)$ (on-top), and $(2/3, 1/3)$ (hcp hollow). In addition to a completely bulk-like structure, we therefore tested various models with adatoms (both oxygen or metal atoms) located at some or all of these allowed locations and free to move within one unit cell away from the surface \cite{Valtiner:2010p3053}.

The results of these survey fits draw a consistent picture of the surface or interface structure across all samples, including those with metal layers. The occupation parameters of Zn reveal a sharp surface/interface with only one partially occupied atomic layer, where $p_{\mathrm{occ}}(\mathrm{Zn})<40$\%. The corresponding oxygen occupation within this DL is, however, close to unity, as is the Zn occupation one DL below. Additionally, we usually observe a comparable partial occupation of oxygen atoms in the next DL above the incomplete Zn-layer in the on-top position, while the occupations of the corresponding Zn atom and any adatoms in fcc and hcp hollow sites are negligible in that layer. Hence, there is strong evidence for an oxygen overlayer on-top of the nominally Zn-terminated surface, while there are no indications for the presence of any other ordered adsorbed species or, in particular, the ordering of metal atoms in those samples with metal Schottky contacts. Apart from the oxygen overlayer atoms, which show an outward relaxation of approximately 50 pm, there are no significant movements ($\Delta z < 10$pm), even in the topmost atomic layers.\cite{Jedrecy:2000p2901,Wander:2001p2905,Valtiner:2010p3048} The DW factors converge to values close to those reported for bulk crystals \cite{Sawada:1996ho} and manually modifying them has no significant influence on the optimized values of any other parameters.

\begin{figure}
  \includegraphics[width=1.0\columnwidth]{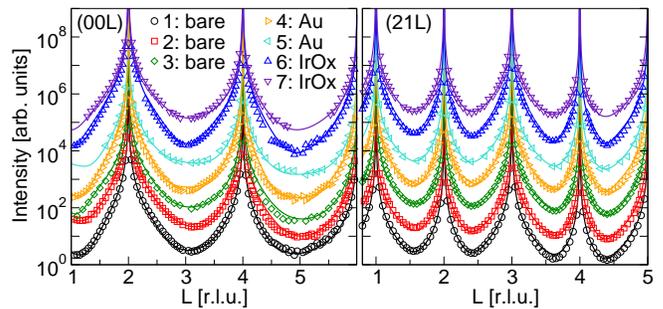}
  \caption{\label{fig:roddata}(color online) Measured diffraction data (open symbols) and corresponding calculated intensities (lines) for two representative CTRs on all samples. Error bars are smaller than the data points and have been omitted.}
\end{figure}

For a direct comparison of all samples, the trends identified in the survey fits were captured in one model, which was systematically applied to all data sets. It consists of a bulk-like ZnO unit cell with an extra oxygen overlayer added on top of the terminating Zn layer, and includes occupation parameters, DW factors, and $z$-displacements for the two topmost DLs (Zn1, O1, Zn2, O2; see Fig.~\ref{fig:ZnO_structure}) and the overlayer (O$_{\mathrm{OL}}$). This model yields good fits on both bare and metal coated surfaces. Fig.~\ref{fig:roddata} shows the excellent agreement between measured diffraction data and simulated CTR profiles for all samples, and the corresponding final FOM values are summarized in Table~\ref{tab:results}.

\begin{figure}
  \includegraphics[width=1.0\columnwidth]{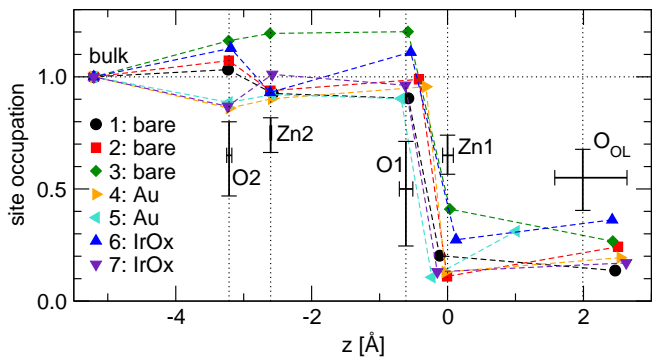}
  \caption{\label{fig:fitresults}(color online) Fitted $z$-positions and site occupations for all samples. Average error bars for the parameter values are shown next to the atom labels. These are defined by a 5\% increase in the FOM value. Dotted vertical lines mark the nominal bulk positions.}
\end{figure}

Fitted $z$-positions and occupation parameters are plotted in Fig.~\ref{fig:fitresults}. The presence of the oxygen overlayer is clearly seen, as the occupation of the two topmost oxygen layers (O$_{\mathrm{OL}}$, O1) is equal to those of the corresponding zinc layers immediately below (Zn1, Zn2). The low occupation ($<$ 35\%) of Zn1 and O$_{\mathrm{OL}}$ is most likely attributed to the presence of small islands on an otherwise atomically flat surface. Thus, a large fraction ($>$ 65\%) of the O1 atoms form the overlayer on top of the fully occupied Zn2 layer where the islands are absent. No significant atomic movements within the ZnO are observed (Zn2, O2: $\Delta z < 5$pm), but the tendency for outward displacements of the O1 and O$_{\mathrm{OL}}$ atoms suggests an overlayer bond length larger than the Zn--O bond length in the bulk. \footnote{The particular error bar for $z$(O$_{\mathrm{OL}}$) of sample 5 is very large and encompasses the determined positions for the other samples.}

\begin{table}
\caption{\label{tab:results}Final $R_{\mathrm{log}}$ and $R$ values on all samples [$\times 10^{-2}$].}
\begin{ruledtabular}
\begin{tabular}{lccccccc}
  Sample No.& 1 & 2 & 3 & 4 & 5 & 6 & 7\\ \hline
  $R_{\mathrm{log}}$
      & 3.87 & 1.54 & 1.42 & 1.75 & 1.41 & 2.03 & 1.49\\
  $R$
      & 6.22 & 6.37 & 6.71 & 7.56 & 4.67 & 8.25 & 4.11 \\
\end{tabular}
\end{ruledtabular}
\end{table}


The (1$\times$1) oxygen overlayer on the bare Zn-polar ZnO surface is most likely in the form of hydroxyl groups (OH), previously observed by x-ray photoemission spectroscopy (XPS),\cite{Valtiner:2007p3049,Allen:2011a} although we lack sufficient sensitivity to detect hydrogen with SXRD. XPS measurements on the same hydrothermal ZnO material indicate a stable OH coverage of at least 1 monolayer (ML) on the Zn-polar face.\cite{Allen:2011a} However, the formation of this fully occupied (1$\times$1) overlayer is not well understood, as such an arrangement does not obey the electron counting rule, and is predicted to be thermodynamically unfavorable by density-functional theory (DFT).\cite{Wander:2001p2898,Meyer:2004p3051} Furthermore, this fully occupied overlayer does not appear in any of the calculated phase diagrams for ZnO, which predict no stable phases with greater than 1/2 ML OH coverage and OH adsorption at fcc-hollow rather than on-top sites.\cite{Valtiner:2010p3048,Valtiner:2009p2890,Kresse:2003p2990} Our results therefore indicate that alternative stabilization mechanisms may play a role at surfaces prepared under typical device fabrication conditions.

Interestingly, the ordered (1$\times$1) oxygen layer remains intact when the surfaces are covered with plain or oxidized metal Schottky contacts.  This is consistent with \emph{ab initio} calculations of the adsorption of Cu atoms on polar ZnO surfaces and the role of chemical bonding at metal-ZnO interfaces, which indicate that metal-zinc bonding is unfavorable and associated with ohmic rather than Schottky contact behavior.\cite{Hegemann:2008gb,Dong:2007ku} The question as to whether metal adsorption occurs on top of the surface OH groups or by replacing the H atoms remains open, as the detection of hydrogen at buried interfaces is extremely challenging.

In conclusion, SXRD showed that the atomic structure of the bare ZnO (0001) Zn-polar surface prepared under typical device fabrication conditions has a bulk-like termination with no significant atomic relaxations. Most interestingly, a (1$\times$1) overlayer of oxygen atoms on top of the terminating zinc atoms was observed, consistent with XPS measurements, but at odds with DFT calculations. At bare surfaces, this (1$\times$1) oxygen overlayer is most likely associated with the presence of hydroxyl (OH) groups. Significantly, the (1$\times$1) oxygen overlayer remains in place following the fabrication of both plain and oxidized metal Schottky contacts.


\begin{acknowledgments}
The authors thank Matts Bj{\"o}rck for supporting \GenX\ and making it freely available.\footnote{M. Bj{\"o}rck, \GenX\ homepage, http://genx.sourceforge.net} Excellent beamline support by P.R. Willmott, P.J. Eng, and the staff of the SLS and APS is gratefully acknowledged. Use of the Advanced Photon Source, an Office of Science User Facility operated for the U.S. Department of Energy (DOE) Office of Science by Argonne National Laboratory, was supported by the U.S. DOE under Contract No. DE-AC02-06CH11357.
This work was supported by the New Zealand Marsden Fund (Contract No. UOC0909), and the U.S. Department of Energy (Contract No. DE-FG02-06ER46273).
\end{acknowledgments}


\bibliography{ZnO_Zn}

\end{document}